\documentstyle[psfig]{kapproc}  
\let\footnote\savefootnote

\normallatexbib

\newcommand{\EQ}{\begin{equation}}
\newcommand{\EN}{\end{equation}}
\newcommand{\bea}{\begin{eqnarray}}
\newcommand{\eea}{\end{eqnarray}}

\begin{document}

\articletitle[Field Theory of Scaling Lattice Models]
{Field theory of scaling \\
lattice models.\\
The Potts antiferromagnet\footnote{Talk given at the NATO Advanced Research 
Workshop on Statistical Field Theories, Como 18-23 June 2001}}

\author{Gesualdo Delfino}

%% affil, email, and abstract are optional
\affil{SISSA\\
via Beirut 2-4, 34014 Trieste\\
Italy}
\email{delfino@sissa.it}

%% optional, to supply a shorter version of the title for the running head:
%%\chaptitlerunninghead{}

\begin{abstract}
In contrast to what happens for ferromagnets, the lattice structure 
participates in a crucial way to determine existence and type of critical 
behaviour in antiferromagnetic systems. It is an interesting question to
investigate how the memory of the lattice survives in the field theory
describing a scaling antiferromagnet. We discuss this issue for the square 
lattice three-state Potts model, whose scaling limit as $T\rightarrow 0$ is
argued to be described exactly by the sine-Gordon field theory at a specific
value of the coupling. The solution 
of the scaling ferromagnetic case is recalled for comparison. The field theory
describing the crossover from antiferromagnetic to ferromagnetic behaviour
is also introduced.
\end{abstract}


\section{Introduction}

The description of statistical systems nearby their phase transition points
is one of the most stimulating applications of quantum field theory. 
The typical problem one has to face in this context is that of 
{\em non-trivial} fixed points of the renormalisation group, i.e. of {\em 
strongly interacting} quantum field theories. It is then an important 
discovery of the last 
years that quantum field theory can actually be done non-perturbatively in two 
dimensions. Here, conformal \cite{BPZ} and integrable \cite{Taniguchi} field 
theories describe {\em exactly} the critical points and the scaling limits of 
statistical models, respectively. 

This fact has been exploited to solve the scaling ferromagnetic models in 
two dimensions directly in the continuum limit. 
Since scaling ferromagnets exhibit universal 
behaviour independent on lattice details, the recipe is\,: find the simplest 
integrable field theory with the internal symmetry characteristic of the 
given universality class. It is remarkable that the basic two-dimensional 
ferromagnets 
%(the Ising model in a 
%magnetic field \cite{Taniguchi}, the $q$-state Potts model \cite{qPotts}, 
%the $O(n)$ model \cite{Oenne}, just to make few examples) 
are integrable in the 
continuum limit (not on the lattice, usually) and can be solved in this way.

It should not be surprising that much less is known for
antiferromagnets. The spins in an antiferromagnet want to be in a state which
is different from that of their nearest neighbours. Hence, the number of such
neighbours, namely the lattice structure, participates in a crucial way to 
determine existence and type of critical behaviour. It follows that the 
phenomenology is much richer than in the ferromagnetic case. Nevertheless, if 
a critical point with infinite correlation length exists, there must be a 
quantum field theory describing the scaling limit. It is an intriguing 
question to find such a theory and to understand how it `remembers' about
the lattice structure. Once the lattice and internal symmetries have been 
disentangled in the field theory of the scaling antiferromagnet, another 
interesting point is that of breaking the first symmetry while preserving the 
second. The result should be a theory describing the crossover from
antiferromagnetic to ferromagnetic behaviour.

Once again, one expects that in two dimensions these issues can be investigated
in a precise, non-perturbative way. We will show that this is indeed the case 
for a non-trivial example, the three-state Potts 
model on the square lattice. In the next Section we recall the solution of 
the ferromagnetic case before turning to the antiferromagnet in Section 3 and
discussing the crossover between the two in Section 4.

\section{Scaling Limit of the Ferromagnetic Three-state Potts Model}

The three-state Potts model is defined by the Hamiltonian
\EQ
H=-J\sum_{\langle i,j\rangle}\delta_{s_i,s_j}\,\,,
\hspace{1cm}s_i=1,2,3
\label{lattice}
\EN
where $s_i$ denotes the spin located at the $i$-th site of a regular lattice 
and the sum is taken
over nearest neighbours. The model is characterised by the invariance under
global permutations of the values of the spin. The permutation group $S_3$ 
can be seen as the product of the group $Z_3$ of cyclic permutations times
a ``charge conjugation'' $C$. The elementary $Z_3$ transformation and 
charge conjugation act as follows on the complex spin variable 
$\sigma_j=e^{2i\pi s_j/3}$\,:
$$
Z_3\,:\hspace{.3cm}\sigma_j\rightarrow e^{2i\pi/3}\,\sigma_j\,,\hspace{1.5cm}
C\,:\hspace{.3cm}\sigma_j\rightarrow\sigma_j^*\,\,;
$$
of course they leave invariant the energy operator 
$\varepsilon_j=\sum_i\delta_{s_i,s_j}$.

The model exhibits ferromagnetic or antiferromagnetic behaviour depending on
the sign of the coupling $J$. 

In two dimensions, the ferromagnetic ($J>0$) model undergoes a second order
phase transition at a critical temperature $T_c$ \cite{Baxter}. It was shown 
in \cite{Dotsenko,FZ} that the critical point is described by the minimal model
of conformal field theory with central charge $C=4/5$. The spin 
operator $\sigma(x)$
and the energy operator $\varepsilon(x)$ correspond to the primary
conformal operators $\phi_{2,3}(x)$ and $\phi_{2,1}(x)$ with scaling dimensions
$X_\sigma=2/15$ and $X_\varepsilon=4/5$, respectively. As a consequence,
the scaling limit can be described by adding to the fixed point action 
${\cal A}_{C=4/5}$ the thermal perturbation in the form
\EQ
{\cal A}_F={\cal A}_{C=4/5}+\tau\int d^2x\,\varepsilon(x)\,,
\label{action}
\EN
with $\tau$ measuring the deviation from $T_c$. While the model (\ref{lattice})
is not solvable on the lattice away from $T=T_c$ \cite{Baxter}, the scaling
limit (\ref{action}) belongs to the large class of 
integrable quantum field theories discovered by A. Zamolodchikov 
\cite{Taniguchi}. 

Integrable quantum field theories can be solved exactly in the scattering
theory framework \cite{ZZ}. In fact, integrability (i.e. the existence of an 
infinite number of quantum integrals of motion) ensures the complete elasticity
and factorisation of the scattering processes and allows the determination
of the scattering amplitudes. Universality requires that the scaling limit
of the ferromagnetic three-state Potts model corresponds to the simplest
integrable scattering theory implementing the $S_3$ symmetry. The difference
between the high and low temperature phases is made by the nature of the 
excitations. 

At $T>T_c$ there is a single ground state and the simplest realisation of the 
symmetry is in terms of a doublet of charge conjugated particles $A$ and 
$\bar{A}$ of mass $m$ transforming under $S_3$ as the spin operators $\sigma$ 
and $\sigma^*$, respectively. The $Z_3$ symmetry in enforced by requiring 
the existence of the fusion process
\EQ
A\,A\rightarrow \bar{A}\,\,.
\label{fusion}
\EN
Factorisation of multiparticle processes implies that the full scattering 
matrix is determined by the three two-particle amplitudes depicted in Fig.\,1.
The last process turns out to be incompatible with factorisation and 
the property (\ref{fusion}) together, and the simplest solution with 
$S_3^F(\theta)=0$ is \cite{KS,Sasha}
\EQ
S_1^F(\theta)\,=\,S_2^F(i\pi-\theta)\,\,=\,\,
\frac{\sinh\frac12(\theta+\frac{2i\pi}{3})}
                           {\sinh\frac12(\theta-\frac{2i\pi}{3})}\,,
\EN
where $\theta$ parameterises the center of mass energy
$\sqrt{s}=2m\cosh(\theta/2)$.
The nonvanishing amplitudes are related by crossing symmetry while unitarity
follows from the fact that complex conjugation amounts to $\theta
\rightarrow -\theta$.
The pole at $\theta=2\pi i/3$ in the amplitude $S_1^F(\theta)$ corresponds to 
the bound state (\ref{fusion}).

At $T<T_c$ there are three degenerate ground states and the excitations are 
``kinks'' $K_{j,j\pm 1}$ interpolating between the ground state $j=1,2,3$ 
and the ground state $j+1$ (mod 3). Their space-time trajctories draw domain
walls separating regions with different magnetisation. Due to invariance 
under permutations, there are only three inequivalent two-kink scattering 
amplitudes which are readily mapped into those of Fig.\,1 through the 
identifications $K_{j,j+1}\longleftrightarrow A$, 
$K_{j,j-1}\longleftrightarrow \bar{A}$. As a consequence the solution for the
scattering amplitudes is the same than at $T>T_c$. 
This is the way in which the high-low temperature duality of the Potts model
emerges in this context. The computation of the correlation functions in the 
two phases starting from the scattering solution can be found in \cite{DC}.

\begin{figure}
\centerline{
\psfig{figure=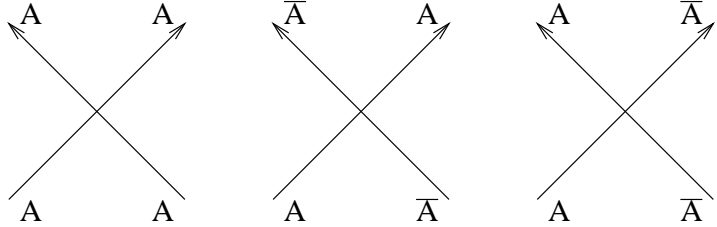}}
\caption{The scattering amplitudes $S_1$, $S_2$ and $S_3$.} 
\end{figure}

\section{The Square Lattice Antiferromagnet}
At $T=0$ the antiferromagnetic ($J<0$) three-state Potts model on the {\em 
square lattice} finds infinitely many ground states in which each spin
has a value different from that of its nearest neighbours. 
After transferring the labels from sites to faces, a configuration
in terms of arrows can be obtained through the following rule: if an 
observer in one face labelled $j$ looks across an edge to an adjacent face
labelled $j+1$ (mod 3), then put an arrow on this edge pointing to the 
observer's left; if the adjacent face is labelled $j-1$ (mod 3), point the 
arrow to the right. All sites of the resulting configuration 
satisfy the rule ``two arrows in, two arrows out'' which defines the 
six-vertex model. The latter is exactly solvable on the lattice \cite{Baxter}
and is known to be critical and equivalent at long distance to a free massless 
boson. 

This exact mapping of the square lattice antiferromagnet (\ref{lattice}) 
at zero-temperature onto a particular case\footnote{The one in which all 
vertices are equally weighted, the so called ice point.} of the six-vertex 
model has been known for long time and has been used to determine the scaling
dimensions of a number of relevant operators \cite{dNNS,PW,BH,SS}. These
are the staggered magnetisation 
$\Sigma_j=(-1)^{j_1+j_2}\,e^{2i\pi s_j/3}$ with dimension $1/6$, 
the uniform magnetisation $\sigma_j=e^{2i\pi s_j/3}$ with
dimension $2/3$, and the staggered polarisation\footnote{The primed sum 
indicates summation over the next nearest neighbours of $j$.} 
${\cal P}_j=(-1)^{j_1+j_2}\sum'_i(2\delta_{s_i,s_j}-1)$ with 
dimension $3/2$. Here and in the following we identify the $j$-th site of the
square lattice through a pair of integers $(j_1,j_2)$, and call even (odd)
sublattice the collection of the sites with $j_1+j_2$ even (odd). 

The model is not solved on the lattice at non-zero temperature and the 
issue of the approach to criticality (which involves first of all the 
determination of the scaling dimension of the energy operator 
$\varepsilon_j=\sum_i\delta_{s_i,s_j}$) has made the object of recent studies
\cite{FS,CJS,af}. The authors of \cite{CJS} exploited a mapping onto a height
model to study the lattice excitations at $T>0$ and explain the anomalous
corrections to scaling observed in simulations. Here we will follow
Ref. \cite{af} to show how the scaling limit as $T\rightarrow 0$ is in fact
described by an integrable quantum field theory.

Since the critical model is described by a gaussian fixed point, the 
expectation that the scaling limit corresponds to a massive integrable
field theory is very natural. In fact, under the (mild) assumption that
the thermal operator $\varepsilon_j$ gives in the continuum limit a single
{\em relevant} operator $\varepsilon(x)$, the action one is left with for the
scaling limit is that of the sine-Gordon model,
\EQ
{\cal A}_{AF}=\int d^2x\,\left(\frac12\,\partial_a\varphi\partial^a\varphi
-\mu\cos\beta\varphi\right)\,,
\label{sg}
\EN
which is integrable. Having said that, it remains to be understood how the 
action (\ref{sg}) actually describes the Potts antiferromagnet, namely where 
this action hides the relevant lattice and $S_3$ symmetries, how the Potts 
degrees of freedom
are expressed in terms of the bosonic field $\varphi$, and which value of 
$\beta$ determines the scaling dimension $X_\varepsilon=\beta^2/4\pi$ 
of the energy operator $\varepsilon=\cos\beta\varphi$.

To answer these questions we have to recall few facts about the operator 
content of the sine-Gordon model. At the gaussian fixed point ($\mu=0$)
the bosonic field decomposes into holomorphic and antiholomorphic parts as
$\varphi(x)=\phi(z)+\bar{\phi}(\bar{z})$, 
where $z=x_1+ix_2$ and $\bar{z}=x_1-ix_2$. The scaling operators 
$V_{p,\bar{p}}(x)=\exp{i[p\,\phi(z)+\bar{p}\bar{\phi}(\bar{z})]}$
have scaling dimension $X=(p^2+\bar{p}^2)/8\pi$, spin $s=(p^2-\bar{p}^2)/8\pi$,
and satisfy the gaussian operator product expansion
\EQ
V_{p_1,\bar{p}_1}(x)V_{p_2,\bar{p}_2}(0)=
z^{p_1p_2/{4\pi}}\,\bar{z}^{\bar{p}_1\bar{p}_2/{4\pi}}\,
V_{p_1+p_2,\bar{p}_1+\bar{p}_2}(0)+\ldots\,.
\label{ope}
\EN
This relation shows that taking $V_{p_1,\bar{p}_1}(x)$ around 
$V_{p_2,\bar{p}_2}(0)$ by sending $z\rightarrow ze^{2i\pi}$ and 
$\bar{z}\rightarrow \bar{z}e^{-2i\pi}$ produces a phase factor
$e^{2i\pi\gamma_{1,2}}$, where
$\gamma_{1,2}=(p_1p_2-\bar{p}_1\bar{p}_2)/(4\pi)$
is called index of mutual locality. If $\gamma_{1,2}$ is an integer the 
correlators $\langle ..V_{p_1,\bar{p}_1}(x)V_{p_2,\bar{p}_2}(0)..\rangle$ are
single valued as functions of $x$ and the two operators are said to be 
mutually local. Since $\gamma_{1,1}=2s$, the operators which are local with 
respect to themselves must have integer or half integer spin.

The operators of interest for the description of the statistical model
are scalar ($s=0$) and local with respect to the energy 
$\varepsilon=\cos\beta\varphi$. These requirements select
\bea
V_p&\equiv& V_{p,p}=\exp [ip\varphi]\,,
\label{Vp}\\
U_m &\equiv& V_{2\pi m/\beta,-2\pi m/\beta}=
\exp [2i\pi m\tilde{\varphi}/\beta]\,,\hspace{.3cm}m=\pm 1,\pm 2,..
\label{Um}
\eea
where $\tilde{\varphi}(x)\equiv\phi(z)-\bar{\phi}(\bar{z})$ is sometimes 
called the ``dual'' boson. A slightly more general analysis extended to the
spin 1/2 operators shows that the integer $m$ in (\ref{Um}) is in fact the
topologic charge that in the sine-Gordon model originates from the periodicity 
of the potential. Hence the operators $U_m(x)$ have charge $m$ while the 
operators $V_p(x)$ are neutral. This implies in particular that in the model
(\ref{sg})
\EQ
\langle V_p\rangle\neq 0\,,\hspace{1cm}\langle U_m\rangle=0\,\,.
\label{vevs}
\EN
Another point to be remarked is that the operators $U_m(x)$ with $|m|>3$ are 
always irrelevant ($X>2$) as long as the perturbation in (\ref{sg}) is
relevant (i.e. $\beta^2<8\pi$).

The lattice operators $\Sigma_j$, $\sigma_j$ and ${\cal P}_j$ are not 
invariant under the $S_3$ symmetry and/or the exchange of the even and odd
sublattices. Hence their continuum counterparts have to be sought among the 
$U_m(x)$. Comparison with the known scaling dimensions shows that the 
matching is complete provided we take $\beta=\sqrt{6\pi}$, 
what in turn implies $X_\varepsilon=3/2$. The operator identifications are 
summarised in Table\,1 from which the following correspondences can also be 
read:
\begin{center}
$Z_3$ charge =\,\,$m$\,(mod 3)

$C\,=$ complex conjugation

sublattice parity = $(-1)^m$\,\,.
\end{center}
\noindent
Hence we see that in the continuum limit both the $Z_3$ symmetry
and the lattice symmetry are ruled by the topologic charge $m$.

\begin{table}[ht]
\caption{Relevant operators on the lattice and their 
continuum counterparts in the sine-Gordon model.}       
\begin{tabular}{c c l c c}\sphline
 & Lattice definition & Continuum limit & $X$ & m \\ \sphline
$\Sigma$ & $(-1)^{j_1+j_2}\,\exp[2i\pi s_j/3]$ & $U_1=\exp[i\sqrt{2\pi/3}\,
\tilde{\varphi}]$ & $1/6$ & $1$ \\ 
$\sigma$ & $\exp [2i\pi s_j/3]$ & $U_{-2}=
\exp [-i\sqrt{8\pi/3}\,\tilde{\varphi}]$ & $2/3$ & $-2$ \\
${\cal P}$ & $(-1)^{j_1+j_2}\sum_i'(2\delta_{s_i,s_j}-1)$ &
$U_3+U_{-3}=\cos\sqrt{6\pi}\,\tilde{\varphi}$ & $3/2$ & $\pm 3$ \\ 
$\varepsilon$ & $\sum_i\delta_{s_i,s_j}$ & $V_{\sqrt{6\pi}}+V_{-\sqrt{6\pi}}=
\cos\sqrt{6\pi}\,\varphi$ & $3/2$ & $0$ \\
\sphline
\end{tabular}
\end{table}
                                 
Due to the integrability of the sine-Gordon model, also the scaling 
antiferromagnet admits an exact scattering description. This time the 
elementary excitations are the soliton $A$ and antisoliton $\bar{A}$ 
interpolating between adjacent sine-Gordon vacua. They carry topologic
charge 1 and $-1$, respectively, and then transform under the symmetries as 
the {\em staggered} magnetisation $\Sigma$ and $\Sigma^*$. This is what makes 
the difference with the ferromagnetic case at the level of the scattering 
theory: the sublattice parity (which plays no role in the ferromagnet) 
would now be violated by the fusion process (\ref{fusion}), which is therefore
forbidden. One consequence is that the last amplitude in Fig.\,1 is no longer 
forced to vanish. The three amplitudes are the sine-Gordon ones \cite{ZZ},
\bea
S_1^{AF}(\theta)= S_2^{AF}(i\pi-\theta)=
-\exp\left\{\int_0^\infty\frac{dx}{x}\frac{\sinh\frac{x}{2}\left(1
-\frac{\xi}{\pi}\right)}{i\sinh\frac{x\xi}{2\pi}\cosh\frac{x}{2}}
\sin\frac{\theta x}{\pi}\right\}\\
S_3^{AF}(\theta)= -\frac{\sinh\frac{i\pi^2}{\xi}}
               {\sinh\frac{\pi}{\xi}(\theta-i\pi)}\,\,S_1^{AF}(\theta)\,,
\hspace{4.8cm}
\eea
evaluated at the value $\xi=3\pi$ corresponding to $\beta=\sqrt{6\pi}$. This
value falls in the sine-Gordon
repulsive region in which the solitons do not form any bound state. In 
particular no asymptotic particle corresponding to the field $\varphi$ 
in (\ref{sg}) is present in the spectrum.

Correlation functions can be computed starting from the scattering theory
through the form factor approach (see \cite{af}). Here we only mention few 
straighforward predictions for the antiferromagnet dictated by the topologic 
charge of the operators. Defining the `exponential' correlation 
length $\xi_\Phi$ associated to an operator $\Phi(x)$ as
\EQ
\langle\Phi(x)\Phi^*(0)\rangle_{\mbox{connected}}\sim\exp(-|x|/\xi_\Phi)\,,\hspace{1cm}
|x|\rightarrow\infty\,,
\EN
then the following universal ratios should be observed in simulations as
$T\rightarrow 0$\,:
$\xi_\sigma/\xi_\Sigma=\xi_{\cal E}/\xi_\Sigma=1/2$, $\xi_{\cal P}/
\xi_\Sigma=1/3$.

\section{Crossover from Antiferromagnetic to Ferromagnetic Behaviour}
We see from Table\,1 that the field theory of the scaling antiferromagnet 
contains a {\em single} relevant operator with the symmetry properties 
required to break the sublattice symmetry while keeping the $S_3$ symmetry.
Therefore the action that should describe the crossover to ferromagnetic 
behaviour reads
\bea
{\cal A}_{cross} &=& {\cal A}_{AF}-\tilde{\mu}\int d^2x\,{\cal P}(x)\nonumber\\
         &=& {\cal A}_{C=1}-\int d^2x\,[\,
\mu\cos\sqrt{6\pi}\,\varphi+\tilde{\mu}\cos\sqrt{6\pi}\,\tilde{\varphi}\,]\,,
\label{cross}
\eea
and defines a one-parameter family of renormalisation group trajectories
(labelled by $\mu/\tilde{\mu}$)
flowing out of the gaussian ($C=1$) fixed point. Since ${\cal P}(x)$ has 
topologic charge $\pm 3$, it reintroduces in the theory the three-particle 
vertex (\ref{fusion}), so that the $Z_3$ symmetry is again manifest.

One of the trajectories described by (\ref{cross}) should flow into the 
ferromagnetic ($C=4/5$) fixed point in the infrared limit. This trajectory
marks a phase boundary across which a continous ordering phase transition 
takes place. The existence of such a transition can be argued as follows. 
When $\tilde{\mu}=0$ we are in the sine-Gordon model (\ref{sg}) in which 
the vacuum expectation value of the spin operator
$\sigma=\exp[-i\sqrt{8\pi/3}\,\tilde{\varphi}]$ (i.e. the spontaneous
magnetisation) vanishes according to (\ref{vevs}). It can be argued in the 
spirit of Ref.\,\cite{msg} that, since ${\cal P}(x)$ is local with respect 
to the solitons of the theory (\ref{sg}), no phase transition takes place
as soon as the perturbation is switched on, namely $\langle\sigma\rangle=0$
for $\tilde{\mu}\simeq 0$. On the other hand, we can perform similar 
considerations for the opposite limit ($\mu\simeq 0$) of the action
(\ref{cross}), where we are perturbing around a sine-Gordon model of the 
{\em dual} boson $\tilde{\varphi}(x)$. The operator $\sigma(x)$ has zero 
topologic charge with respect to this model and its vacuum expectation value
no longer vanishes, namely $\langle\sigma\rangle\neq 0$ for $\mu\simeq 0$.
These conclusions for the two limits are compatible if a phase transition 
takes place in between giving rise to the phase diagram of Fig.\,2.

\begin{figure}
\centerline{
\psfig{figure=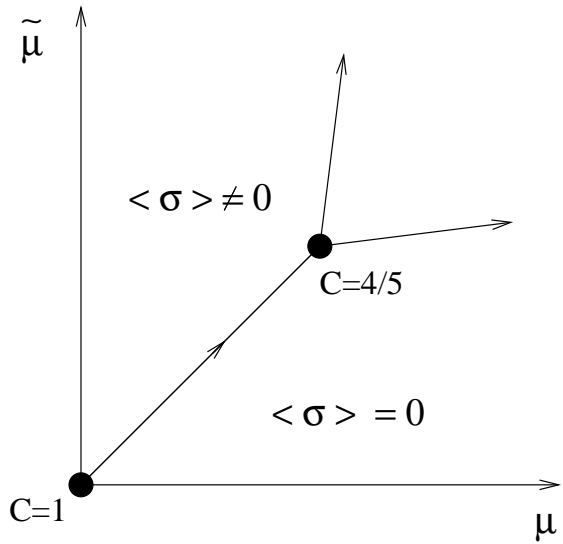}}
\caption{Phase diagram associated to the action 
(\ref{cross}).}
\end{figure}

Generically, the theory (\ref{cross}) will not be integrable when both 
perturbations are switched on. There are reasons to expect, however, that
the massless trajectory connecting the two fixed points is integrable.
As a matter of fact, some years ago Fateev and Al. Zamolodchikov used the 
thermodynamic Bethe ansatz to provide evidence for an infinite series of 
integrable flows between the $Z_N$ parafermionic conformal field theories 
(with central charge $2(N-1)/(N+2)$) and the unitary minimal models ${\cal
M}_{N+1}$ (central charge $1-6/[(N+1)(N+2)]$) \cite{FalZ}. The case $N=4$ 
gives a flow between the fixed points with central charges $C=1$ and $C=4/5$ 
which should correspond to the one we are discussing in connection with the 
Potts model and the action (\ref{cross}). 

A relation between $Z_4$ symmetry and the Potts antiferromagnetic fixed point 
was suggested in Ref. \cite{Saleur}. It should be stressed, however, 
that this symmetry corresponds to an universality class which differs from 
that of the Potts antiferromagnet\footnote{The scaling dimensions of the 
spin and energy operators in the $Z_4$ model are 1/8 and 2/3, respectively 
\cite{FZ}. The possibility that the ``same'' quantum field theory 
accounts for different universality classes is well known (see e.g. Ref. 
\cite{rsos}, Section 6, for a discussion and references on the argument).},
and that it is explicitely broken in the off-critical actions (\ref{sg}) and 
(\ref{cross}). According to the discussion of this section,
the flow from $C=1$ to $C=4/5$ should be described by a factorised massless 
$S$-matrix exhibiting the $S_3$ symmetry characteristic of the 3-state Potts
model. To the best of our knowledge this $S$-matrix did not appear in the 
literature so far.

\section{Conclusion}
We have seen that both the ferromagnetic and (square lattice) antiferromagnetic
scaling limits of the three-state Potts model correspond to integrable 
quantum field theories, and that the massless crossover trajectory should also 
be integrable. It would be interesting to confirm this latter point by 
determining the exact massless $S$-matrix. Of course, one can think of 
extending the investigation of this paper to other antiferromagnets 
possessing a critical point.

We conclude by mentioning that actions of the type (\ref{cross})
containing both kinds of scaling operators (\ref{Vp}) and (\ref{Um}) are not
uncommon in the description of scaling lattice models. For example, the 
Ashkin-Teller model consists of two Ising models coupled through their 
energy terms (four spin interaction). As long as the two Ising models are 
kept at the same temperature the scaling limit is described 
by the 
action (\ref{sg}) with $\beta$ parameterising this time the line of fixed
points characteristic of the Ashkin-Teller model (see \cite{AT}). A 
temperature difference corresponds to the addition of the operator
$\cos[4\pi\tilde{\varphi}/\beta]$. It can be argued that also this action
admits a massless flow which in this case ends in the infrared limit
into an Ising fixed point with $C=1/2$.

\vspace{.3cm}
\noindent
{\bf Acknowledgments:} I thank J. Cardy and V. Fateev for interesting 
discussions.

\notes

\end{document}